\documentclass[preprint,amsmath,amssymb,superscriptaddress]{revtex4-1}

\usepackage{graphicx}
\usepackage{dcolumn}
\usepackage{bm}
\usepackage{color}

\begin{document}


\title{Perpendicular magnetic anisotropy at the Fe/MgAl$_2$O$_4$ interface: Comparative first-principles study with Fe/MgO}


\author{Keisuke Masuda}
\affiliation{Research Center for Magnetic and Spintronic Materials, National Institute for Materials Science (NIMS), 1-2-1 Sengen, Tsukuba 305-0047, Japan}

\author{Yoshio Miura}
\affiliation{Research Center for Magnetic and Spintronic Materials, National Institute for Materials Science (NIMS), 1-2-1 Sengen, Tsukuba 305-0047, Japan}
\affiliation{Electrical Engineering and Electronics, Kyoto Institute of Technology, Kyoto 606-8585, Japan}
\affiliation{Center for Materials Research by Information Integration, National Institute for Materials Science (NIMS), 1-2-1 Sengen, Tsukuba 305-0047, Japan}
\affiliation{Center for Spintronics Research Network (CSRN), Graduate School of Engineering Science, Osaka University, Machikaneyama 1-3, Toyonaka, Osaka 560-8531, Japan}


\date{\today}

\begin{abstract}
We present a theoretical study on interfacial magnetocrystalline anisotropy for Fe/MgAl$_2$O$_4$. This system has a very small lattice mismatch at the interface and therefore is suitable for realizing a fully coherent ferromagnet/oxide interface for magnetic tunnel junctions. On the basis of density functional theory, we calculate the interfacial anisotropy constant $K_{\rm i}$ and show that this system has interfacial perpendicular magnetic anisotropy (PMA) with $K_{\rm i} \approx 1.2\,{\rm mJ/m^2}$, which is a little bit smaller than that of Fe/MgO ($K_{\rm i} \approx$ 1.5--1.7$\,{\rm mJ/m^2}$). Second-order perturbation analysis with respect to the spin-orbit interaction clarifies that the difference in $K_{\rm i}$ between Fe/MgAl$_2$O$_4$ and Fe/MgO originates from the difference in contributions from spin-flip scattering terms at the interface. We propose that the insertion of tungsten layers into the interface of Fe/MgAl$_2$O$_4$ is a promising way to obtain huge interfacial PMA with $K_{\rm i} \gtrsim 3\,{\rm mJ/m^2}$.
\end{abstract}


\maketitle

\section{\label{introduction} introduction}
Perpendicular magnetic anisotropy (PMA) is an essential property for ferromagnets (FMs) in magnetic tunnel junctions (MTJs) to realize nonvolatile magnetic random access memories (MRAMs) \cite{2016Dieny-Wiley}. The PMA is beneficial for obtaining sufficiently high thermal stability and low critical current in spin-transfer-torque MRAMs (STT-MRAMs), in which current-induced spin-transfer torque is used for magnetization switching \cite{2016Dieny-Wiley}. Although large PMA has been observed in several FMs such as $D0_{22}$ ${\rm Mn}_3{\rm Ga}$ \cite{2009Wu-APL,2011Mizukami-APL}, $D0_{22}$ ${\rm Mn}_3{\rm Ge}$ \cite{2012Kurt-APL,2013Mizukami-APEX}, $L1_{0}$ ${\rm MnGa}$ \cite{2011Mizukami-APL}, and $L1_{0}$ ${\rm FePt}$ \cite{1995Klemmer-SMM}, MTJs with these FMs did not show sufficiently high tunnel magnetoresistance (TMR) ratios, which is another important requirement for MRAM applications. Therefore, interfacial PMA at interfaces between FMs and insulator barriers has attracted much attention mainly in MTJs consisting of Fe-based FMs and MgO barriers.

In addition to high TMR ratios \cite{2001Butler-PRB,2001Mathon-PRB,2004Parkin-NatMat,2004Yuasa-NatMat}, interfacial PMA has also been obtained in the MgO-based MTJs. By using thin CoFeB layers ($\sim 1.3\,{\rm nm}$), Ikeda {\it et al.} observed relatively large PMA at the interface of CoFeB/MgO/CoFeB MTJ \cite{2010Ikeda-NatMat}. In subsequent studies \cite{2013Koo-APL,2014Koo-JPD}, Koo {\it et al.} demonstrated that Fe/MgO has a larger interfacial PMA than that of CoFe(B)/MgO, in agreement with theoretical predictions \cite{2011Yang-PRB,2014Zhang-PRB}. Furthermore, interfacial PMA has also been observed in the heterostructure composed of the Heusler alloy Co$_2$FeAl and MgO \cite{2011Wen-APL,2012Wen-APEX}. The interfacial PMA is also advantageous for voltage-torque MRAMs \cite{2016Shiota-APEX}, because high interfacial PMA gives low write error rates in voltage-driven magnetization switching.

The underlying mechanism of such interfacial PMA in Fe-based FM/MgO heterostructures has been discussed in several theoretical studies. By analyzing the local density of states (LDOS) and band structure in Fe/MgO, Nakamura {\it et al.} \cite{2010Nakamura-PRB} clarified that the Fe 3$d_{3z^2-r^2}$ state is distributed away from the Fermi level owing to its hybridization with the O 2$p_z$ state, leading to interfacial PMA. Other studies \cite{2011Yang-PRB,2013Hallal-PRB} also indicated the importance of this hybridization using different theoretical approaches. From a different point of view, the relation between PMA and orbital magnetic moment is another significant issue. A second-order perturbation theory by Bruno \cite{1989Bruno-PRB} revealed a proportional relation between magnetic anisotropy and anisotropy of orbital magnetic moment, which is the so-called Bruno relation. Several theoretical studies have discussed the applicability of the Bruno relation to various Fe-based heterostructures \cite{2013Miura-JAP,2014Zhang-PRB,2017Masuda-PRB_pma}. Moreover, by means of x-ray magnetic circular dichroism (XMCD) measurements, Okabayashi {\it et al.} \cite{2014Okabayashi-APL} showed that the interfacial PMA in Fe/MgO can be explained qualitatively by the Bruno relation. This relation gives valuable information for understanding the interfacial PMA in Fe-based FM/MgO heterostructures.

Although large interfacial PMA has been observed in Fe/MgO heterostructures, the lattice mismatch between Fe and MgO is rather large ($\sim 4\%$), which is a drawback for practical applications. On the other hand, spinel oxide MgAl$_2$O$_4$ has a small lattice mismatch ($<1\%$) with typical FMs such as Fe, Co$_{0.5}$Fe$_{0.5}$, and Co$_2$FeAl$_{0.5}$Si$_{0.5}$ \cite{2010Sukegawa-APL}. Moreover, since the lattice constant of MgAl$_2$O$_4$ can be tuned by changing the Mg/Al composition rate, one can achieve good lattice matching with various FMs. Up to now, relatively high MR ratios have been observed in the MgAl$_2$O$_4$-based MTJs \cite{2010Sukegawa-APL,2012Sukegawa-PRB,2016Belmoubarik-APL,2016Scheike-APEX}. The interfacial PMA has also been obtained in some FM/MgAl$_2$O$_4$ heterostructures \cite{2014Koo-PSS,2014Tao-APL,2017Sukegawa-APL}. In particular, Koo {\it et al.} \cite{2014Koo-PSS} reported that the Fe($0.7\,{\rm nm}$)/MgAl$_2$O$_4$ heterostructure has an interfacial PMA with interfacial anisotropy constant $K_{\rm i}$ of 0.9--1.6$\,{\rm mJ/m^2}$, which is smaller than that of the Fe/MgO heterostructure ($K_{\rm i}\sim$1.5--2.0$\,{\rm mJ/m^2}$) with the same Fe thickness \cite{2013Koo-APL}. Possible reasons for such a difference in $K_{\rm i}$ should be clarified; however, no theoretical study has addressed interfacial magnetocrystalline anisotropy in Fe/MgAl$_2$O$_4$.

In this work, we study interfacial magnetocrystalline anisotropy in Fe/MgAl$_2$O$_4$ by means of first-principles calculations based on density functional theory. We find that this system has interfacial PMA with $K_{\rm i} \approx 1.2\,{\rm mJ/m^2}$. This value of $K_{\rm i}$ is smaller than that calculated in Fe/MgO with a similar barrier thickness ($K_{\rm i} \approx$ 1.5--1.7$\,{\rm mJ/m^2}$), in agreement with the above-mentioned experimental results. To clarify the origin of such a difference in $K_{\rm i}$, second-order perturbation analyses are carried out, which find that the smaller $K_{\rm i}$ in Fe/MgAl$_2$O$_4$ is due to a smaller positive contribution in $K_{\rm i}$ from spin-flip electron scattering. We show that these results can be naturally understood from the features of the LDOSs and band structures in these systems. We finally propose an interfacial insertion of tungsten (W) layers into Fe/MgAl$_2$O$_4$ as a possible way to achieve a larger $K_{\rm i}$. It is shown that such Fe/W/MgAl$_2$O$_4$ systems with 4--5 layers of W have a large $K_{\rm i}$ of $\gtrsim 3\,{\rm mJ/m^2}$.

\section{\label{methods} calculation method}
We analyzed Fe/MgAl$_2$O$_4$(001) and Fe/MgO(001) by means of density functional theory (DFT) including the effect of spin-orbit interactions, which is implemented in the Vienna {\it ab initio} simulation program (VASP) \cite{1996Kresse-PRB}. We adopted the spin-polarized generalized gradient approximation (GGA) \cite{1996Perdew-PRL} for the exchange-correlation energy and used the projector augmented wave (PAW) potential \cite{1994Bloechl-PRB,1999Kresse-PRB} to treat the effect of core electrons properly.

Figures 1(a) and 1(b) show the supercells of Fe(5)/MgAl$_2$O$_4$(9) and Fe(5)/MgO(5) used in this study, where each number in parentheses represents each layer number. Note that MgAl$_2$O$_4$(9) and MgO(5) have similar barrier thicknesses, which are suitable for comparison of interfacial magnetic anisotropy. As mentioned in Sec. \ref{introduction}, the most striking feature of Fe/MgAl$_2$O$_4$ is the significantly small lattice mismatch between the electrode and the barrier; at the interface, two unit cells of bcc Fe with $2\,a_{\rm Fe}=5.732$\,{\AA} can be well fitted to MgAl$_2$O$_4$ with $a_{\rm MgAl_{2}O_{4}}/\sqrt{2}=5.72$\,{\AA}. Thus, we fixed the in-plane lattice constant $a$ of the Fe/MgAl$_2$O$_4$ supercell to $a=2\,a_{\rm Fe}=5.732$\,{\AA}. On the other hand, the lattice mismatch is relatively large in Fe/MgO, for which we used two supercells with different in-plane lattice constants $a$: one is $a=a_{\rm Fe}=2.866$\,{\AA}, and the other is $a=a_{\rm MgO}/\sqrt{2}=2.98$\,{\AA}. In all of these supercells, we carried out structure relaxation, through which optimum atomic positions and the interfacial distance between the electrode and the barrier were determined. Here, we used the known fact that an interfacial atomic configuration where O atoms are on top of Fe atoms [see Figs. 1(a) and 1(b)] is energetically favored in both Fe/MgAl$_2$O$_4$ and Fe/MgO \cite{remark_interface}. The details of our structure relaxation are given in our previous paper \cite{2017Masuda-PRB_bias}.

In each optimized supercell, we calculated interfacial magnetocrystalline anisotropy $K_{\rm i}$ using the well-known force theorem \cite{1990Daalderop-PRB}
\begin{equation}
K_{\rm i}=(E_{[100]}-E_{[001]})/2S,\label{eq:Ki}
\end{equation}
where $E_{[100]}$ ($E_{[001]}$) is the sum of the eigenenergies of the supercell with the magnetization parallel to the $[100]$ ($[001]$) direction, and $S$ is the cross-sectional area of the supercell. Note that the factor 2 in the denominator reflects the fact that each supercell has two interfaces. In order to confirm whether the force theorem gives reliable results for the present systems, we also calculated $K_{\rm i}$ in the self-consistent-field (SCF) manner using the total energies instead of the sum of eigenenergies in Eq. (\ref{eq:Ki}) \cite{1990Daalderop-PRB}. In this paper, we represent a set of $k$-point numbers used for the calculations as $N_{x} \times N_{y} \times N_{z}$, where $N_{x}$, $N_{y}$, and $N_{z}$ are the $k$-point numbers used for the $x$, $y$, and $z$ directions of supercells, respectively. Figure \ref{kdep}(a) shows the values of $K_{\rm i}$ in Fe/MgAl$_2$O$_4$ as a function of the number of in-plane $k$ points $N \equiv N_{x}=N_{y}$ obtained from the force theorem and the SCF total-energy calculation, where $N_{z}$ is fixed to 1 or 3. We see that the value of $K_{\rm i}$ is saturated for $N \gtrsim 19$ in all four of the cases shown in the figure and that the saturated values are almost the same ($K_{\rm i} \sim 1.2\, {\rm mJ/m^2}$). Similar saturations of $K_{\rm i}$ were also obtained in two Fe/MgO systems with $a=a_{\rm MgO}/\sqrt{2}$ and $a=a_{\rm Fe}$, as shown in Figs. \ref{kdep}(b) and \ref{kdep}(c), respectively. In both of these systems, $K_{\rm i}$ is saturated to $\sim 1.6\, {\rm mJ/m^2}$ for $N \gtrsim 37$. All these results indicate that the calculation using the force theorem and $19 \times 19 \times 1$ ($37 \times 37 \times 1$) $k$ points is sufficient to accurately estimate $K_{\rm i}$ in Fe/MgAl$_2$O$_4$ (Fe/MgO) \cite{remark_kpoints}; in the following, we use such calculation conditions. From Eq. (\ref{eq:Ki}), we can easily see that positive (negative) $K_{\rm i}$ indicates the tendency toward perpendicular (in-plane) magnetic anisotropy. However, actual magnetic anisotropy is estimated by $K_{\rm eff}\,t=K_{\rm i}+E_{\rm demag}\,t$, where $t$ is the effective thickness of the Fe electrode and $E_{\rm demag}\,t$ represents magnetic shape anisotropy. The second term $E_{\rm demag}\,t$ always has a negative value, and therefore favors in-plane magnetic anisotropy. In the present work, we calculated $E_{\rm demag}\,t$ by summing up the magnetostatic dipole-dipole interaction between atomic magnetic moments \cite{1990Daalderop-PRB} with the use of the Ewald-summation technique \cite{2000Grzybowski-PRB}.

In addition to these calculations, we further carried out a detailed second-order perturbation analysis to understand magnetocrystalline anisotropy in Fe/MgAl$_2$O$_4$ and Fe/MgO more deeply. By treating the spin-orbit interaction $H_{\rm SO}$ as a perturbation term, the second-order perturbation energy is expressed as
\begin{eqnarray}
  E^{(2)} &=& \sum_{\bm k} \sum^{\rm unocc.}_{n'\sigma'} \sum^{\rm occ.}_{n\sigma} \frac{|\langle {\bm k}n'\sigma' | H_{\rm SO} | {\bm k}n\sigma \rangle|^{2}}{\epsilon^{(0)}_{{\bm k}n\sigma}-\epsilon^{(0)}_{{\bm k}n'\sigma'}}, \label{eq:2ndPT}\\
  H_{\rm SO} &=& \sum_{i} \xi_{i}\, {\bm L}_{i} \cdot {\bm S}_{i},
\end{eqnarray}
where $\epsilon^{(0)}_{{\bm k}n\sigma}$ is the energy of an unperturbed state $| {\bm k}n\sigma \rangle$ with wave vector ${\bm k}$, band index $n$, and spin $\sigma$. The index occ. (unocc.) on the summation means that the sum is over occupied (unoccupied) states of all atoms in the supercell \cite{2013Miura-JPCM,2007Andersson-PRL}. Note here that the state $| {\bm k}n\sigma \rangle$ can be expanded as $| {\bm k}n\sigma \rangle=\sum_{i\mu} c^{{\bm k}n}_{i\mu\sigma} | i\mu\sigma \rangle$, where $\mu$ is an atomic orbital at site $i$ and $c^{{\bm k}n}_{i\mu\sigma}=\langle i\mu \sigma | {\bm k} n \sigma \rangle$ \cite{2013Miura-JPCM}. In the spin-orbit interaction $H_{\rm SO}$, $\xi_{i}$ is its coupling constant at site $i$, and ${\bm L}_i$ (${\bm S}_i$) is the single-electron angular (spin) momentum operator. As the values of $\xi_{i}$, we used $\xi_{\rm Fe}=54.3\,{\rm meV}$, $\xi_{\rm Mg}=47.5\,{\rm meV}$, $\xi_{\rm Al}=10.8\,{\rm meV}$, and $\xi_{\rm O}=24.3\,{\rm meV}$ for Fe, Mg, Al, and O atoms, respectively. The Wigner-Seitz radius of each atom was set to $r_{\rm Fe}=1.302$\,\AA, $r_{\rm Mg}=1.524$\,\AA, $r_{\rm Al}=1.402$\,\AA, $r_{\rm O}=0.820$\,\AA. All these values of spin-orbit coupling constants and Wigner-Seitz radii are those listed in the pseudopotential files in VASP. We used wave functions and eigenenergies obtained in our DFT calculations as unperturbed states and energies in Eq. (\ref{eq:2ndPT}). The magnetocrystalline anisotropy energy within the second-order perturbation $E^{(2)}_{\rm MCA}$ ($\propto K_{\rm i}$) was calculated as $E^{(2)}_{\rm MCA}=E^{(2)}_{[100]}-E^{(2)}_{[001]}$, where $E^{(2)}_{[100]}$ ($E^{(2)}_{[001]}$) is the energy for the magnetization along the [100] ([001]) direction obtained by Eq. (\ref{eq:2ndPT}). In the process of such an analysis, we can decompose $E^{(2)}_{\rm MCA}=\sum_{i} E^{i}_{\rm MCA}$ into four types of terms coming from different electron scattering around the Fermi level:
\begin{equation}
E^{(2)}_{\rm MCA}\!\! =\! \sum_{i} \left( \Delta E^{i}_{\uparrow \Rightarrow \uparrow} + \Delta E^{i}_{\downarrow \Rightarrow \downarrow} + \Delta E^{i}_{\uparrow \Rightarrow \downarrow} + \Delta E^{i}_{\downarrow \Rightarrow \uparrow} \right). \label{eq:E2MCA_decomp}
\end{equation}
Here, $\Delta E^{i}_{\uparrow \Rightarrow \uparrow}$ ($\Delta E^{i}_{\downarrow \Rightarrow \downarrow}$) originates from spin-conserving electron scattering between occupied and unoccupied majority-spin (minority-spin) states. On the other hand, $\Delta E^{i}_{\uparrow \Rightarrow \downarrow}$ ($\Delta E^{i}_{\downarrow \Rightarrow \uparrow}$) corresponds to spin-flip electron scattering from occupied majority-spin (minority-spin) states to unoccupied minority-spin (majority-spin) states. The details of these calculations are given in a previous paper \cite{2013Miura-JPCM}. As we show in the next section, differences in magnetocrystalline anisotropy between different systems can be explained naturally by these second-order perturbation analyses.

\section{\label{resultsdiscussion} results and discussion}
Table \ref{tab1} shows the values of $K_{\rm i}$, $E_{\rm demag}\,t$, $K_{\rm eff}\,t$, $\Delta M_{\rm orb,i}$, and $M_{\rm spin,i}$ for Fe/MgAl$_2$O$_4$ and Fe/MgO obtained in this study. Here, $\Delta M_{\rm orb,i}$ is the anisotropy of the interfacial Fe orbital magnetic moment and $M_{\rm spin,i}$ is the spin magnetic moment at interfacial Fe atoms. We see that Fe/MgAl$_2$O$_4$ has a positive $K_{\rm i}$ of $1.192\,{\rm mJ/m^2}$. Since this value exceeds the negative shape anisotropy ($E_{\rm demag}\,t=-0.895\,{\rm mJ/m^2}$), this system has interfacial PMA ($K_{\rm eff}\,t=0.296\,{\rm mJ/m^2}>0$). Note that O layer is the termination layer of MgAl$_2$O$_4$, as mentioned in Sec. \ref{methods}. Thus, the interfacial hybridization between Fe $3d_{3z^2-r^2}$ and O $2p_z$ states plays a key role for the interfacial PMA of Fe/MgAl$_2$O$_4$ in the same way as Fe/MgO.

The values of $K_{\rm i}$ and $K_{\rm eff}\,t$ in Fe/MgAl$_2$O$_4$ are smaller than those in Fe/MgO. As mentioned in Sec. \ref{introduction}, the relationship between magnetocrystalline anisotropy and anisotropy of the orbital magnetic moment provides important information on PMA in these systems. From Table \ref{tab1}, we find that both $K_{\rm i}$ and $\Delta M_{\rm orb,i}$ of Fe/MgO with $a=a_{\rm MgO}/\sqrt{2}$ are larger than those of Fe/MgAl$_2$O$_4$, which indicates that the Bruno relation ($K_{\rm i} \propto \Delta M_{\rm orb,i}$) holds for these two systems. On the other hand, it seems that this relation is not applicable to Fe/MgO with $a=a_{\rm Fe}$, because this system has a larger $K_{\rm i}$ but a smaller $\Delta M_{\rm orb,i}$ than Fe/MgAl$_2$O$_4$. Therefore, the following second-order perturbation analysis is required to deeply understand interfacial PMA in all of these systems.

In Fig. \ref{pert-ldos_FeMAO}(a), we show the results of the second-order perturbation analysis for the magnetocrystalline anisotropy in Fe/MgAl$_2$O$_4$. We see that the interfacial Fe layer has the largest positive $E^{i}_{\rm MCA}$, which provides the dominant contribution to the positive $K_{\rm i}$ in this system. This indicates that Fe/MgAl$_2$O$_4$ has {\it interfacial} PMA. At the interfacial Fe layer (Fe1), the anisotropy due to minority-spin scattering ($\Delta E^{i}_{\downarrow \Rightarrow \downarrow}$) provides the largest contribution. In order to understand this feature, we utilize the following simplified expressions for the local magnetocrystalline anisotropy \cite{1993Wang-PRB}:
\begin{eqnarray}
E^{i}_{\rm MCA} &\approx& \Delta E^{i}_{\downarrow \Rightarrow \downarrow} + \Delta E^{i}_{\uparrow \Rightarrow \downarrow}, \label{eq:simpEMCA}\\
\Delta E^{i}_{\downarrow \Rightarrow \downarrow} &=& \xi_{i}^2 \sum_{u_{\downarrow},o_{\downarrow}} \frac{|\langle u_{\downarrow} |L^{i}_{z}| o_{\downarrow} \rangle |^{2} - |\langle u_{\downarrow} |L^{i}_{x}| o_{\downarrow} \rangle |^{2}}{\epsilon_{u_{\downarrow}}-\epsilon_{o_{\downarrow}}}, \label{eq:dndn}\\
\Delta E^{i}_{\uparrow \Rightarrow \downarrow} &=& \xi_{i}^2 \sum_{u_{\downarrow},o_{\uparrow}} \frac{|\langle u_{\downarrow} |L^{i}_{x}| o_{\uparrow} \rangle |^{2} - |\langle u_{\downarrow} |L^{i}_{z}| o_{\uparrow} \rangle |^{2}}{\epsilon_{u_{\downarrow}}-\epsilon_{o_{\uparrow}}}, \label{eq:updn}
\end{eqnarray}
where the meanings of $\Delta E^{i}_{\downarrow \Rightarrow \downarrow}$ and $\Delta E^{i}_{\uparrow \Rightarrow \downarrow}$ are the same as those in Eq. (\ref{eq:E2MCA_decomp}). Here, $\xi_i$ is the spin-orbit coupling constant, $L^{i}_{\alpha}\,(\alpha=x,z)$ is the local angular momentum operator at site $i$, and $| o_{\sigma} \rangle$ ($| u_{\sigma} \rangle$) is a local occupied (unoccupied) state with spin $\sigma$ and energy $\epsilon_{o_{\sigma}}$ ($\epsilon_{u_{\sigma}}$). To derive these expressions, it is assumed that the occupied $d$ states in the majority-spin channel are located deep below the Fermi level. Therefore, we neglected $\Delta E^{i}_{\uparrow \Rightarrow \uparrow}$ and $\Delta E^{i}_{\downarrow \Rightarrow \uparrow}$ in Eq. (\ref{eq:simpEMCA}) [compare with Eq. (\ref{eq:E2MCA_decomp})]. Let us now focus on the local density of states (LDOS) of interfacial Fe atoms in Fe/MgAl$_2$O$_4$ shown in Fig. \ref{pert-ldos_FeMAO}(b). From the inset of the figure, we find that the majority-spin states have quite small LDOS around the Fermi level. Thus, we can expect that the term $\Delta E^{i}_{\downarrow \Rightarrow \downarrow}$ provides the dominant contribution in Eq. (\ref{eq:simpEMCA}) in the case of Fe/MgAl$_2$O$_4$, which is consistent with our results shown in Fig. \ref{pert-ldos_FeMAO}(a). In the previous paragraph, we mentioned that the Bruno relation holds in this system, which is reasonable because only $\Delta E^{i}_{\downarrow \Rightarrow \downarrow}$ is taken into account in the derivation of the Bruno relation \cite{1998Laan-JPCM}.

In order to obtain further information on the PMA in Fe/MgAl$_2$O$_4$, we analyzed wave-vector-resolved magnetocrystalline anisotropy and band structures of the supercell. Previous studies using this type of analysis on other ferromagnetic systems have shown that localized $d$ states around the Fermi level provide the dominant contribution to the magnetocrystalline anisotropy \cite{2010Nakamura-PRB,2009Nakamura-PRL,1994Daalderop-PRB,2012Ouazi-NatCom}. Figure \ref{FeMAO_k-resolved}(a) shows the in-plane wave vector (${\bf k}_{\parallel}$) dependence of $\Delta E({\bf k}_{\parallel}) \equiv E_{[100]}({\bf k}_{\parallel})-E_{[001]}({\bf k}_{\parallel})$ \cite{note1}. Here, we plotted the case of $k_{z}=0$ because $k_{z}$ dependence is very weak owing to the long $c$-axis constant of the supercell. Note that $K_{\rm i}$ is proportional to the sum of $\Delta E({\bf k}_{\parallel})$ over all ${\bf k}_{\parallel}$ in the two-dimensional Brillouin zone. We see that large positive anisotropy is obtained around the $\Gamma$ point, which provides the dominant contribution to the interfacial PMA in this system. We can naturally understand this behavior from the band structures of the supercell shown in Figs. \ref{FeMAO_k-resolved}(b) and \ref{FeMAO_k-resolved}(c). Actually, as seen from Fig. \ref{FeMAO_k-resolved}(c), the minority-spin bands around the $\Gamma$ point have $d_{zx}$- and $d_{yz}$-orbital components near the Fermi level, leading to finite values of $\langle d_{yz},\downarrow | L^{i}_{z} | d_{zx},\downarrow \rangle$ and $\langle d_{zx},\downarrow | L^{i}_{z} | d_{yz},\downarrow \rangle$ included in the first term of the numerator of Eq. (\ref{eq:dndn}). Such a band structure is consistent with sharp peaks in the $d_{yz}$- and $d_{zx}$-orbital LDOSs in the minority-spin states around the Fermi level [see Fig. \ref{pert-ldos_FeMAO}(b)].

We next discuss PMA in Fe/MgO to understand the difference from the case of Fe/MgAl$_2$O$_4$. Figure \ref{pert-ldos_FeMgO_a-MgO}(a) shows the results of the second-order perturbation calculations in Fe/MgO with $a=a_{\rm MgO}/\sqrt{2}=2.98$\,{\AA}. In this case, we obtained similar results with Fe/MgAl$_2$O$_4$; anisotropy energy due to minority-spin scattering ($\Delta E^{i}_{\downarrow \Rightarrow \downarrow}$) at the interface provides the dominant contribution to the PMA. The structure of the LDOS at the interfacial Fe atoms is also similar to that of Fe/MgAl$_2$O$_4$ [see Fig. \ref{pert-ldos_FeMgO_a-MgO}(b)]; the majority-spin state has quite small LDOS around the Fermi level. In Fig. \ref{pert-ldos_FeMgO_a-MgO}(a), a non-negligible difference with Fe/MgAl$_2$O$_4$ is that the Fe/MgO has a larger positive spin-flip component $\Delta E^{i}_{\uparrow \Rightarrow \downarrow}$ at the interfacial Fe atoms. To clarify the reason of this behavior, we show the ${\bf k}_{\parallel}$ dependence of $\Delta E({\bf k}_{\parallel})$ in Fig. \ref{FeMgO-aMgO_k-resolved}(a). We find that positive anisotropy occurs mainly around the X point. Similarly to the case of Fe/MgAl$_2$O$_4$, $d_{zx}$ and $d_{yz}$ states in the minority-spin bands give finite values of $\langle L^{i}_{z} \rangle$ as shown in Fig. \ref{FeMgO-aMgO_k-resolved}(c), leading to positive $\Delta E^{i}_{\downarrow \Rightarrow \downarrow}$. In addition, the majority-spin occupied $d_{xy}$ band and minority-spin unoccupied $d_{zx}$ band yield finite values of $\langle d_{zx},\downarrow | L^{i}_{x} | d_{xy},\uparrow \rangle$, as seen from Figs. \ref{FeMgO-aMgO_k-resolved}(b) and \ref{FeMgO-aMgO_k-resolved}(c). This gives positive $\Delta E^{i}_{\uparrow \Rightarrow \downarrow}$ following Eq. (\ref{eq:updn}), which is the reason why this system has the non-negligible contribution from spin-flip scattering in the interfacial PMA.

We also carried out the same perturbation analysis on the other Fe/MgO with $a=a_{\rm Fe}=2.866$\,{\AA}, which provides valuable insight as explained below. Figure \ref{pert-ldos_FeMgO_a-Fe}(a) shows the results of the calculations, in which we find a clear difference from those of Fe/MgAl$_2$O$_4$ and also from those of Fe/MgO with $a=a_{\rm MgO}/\sqrt{2}$. Namely, at the interfacial Fe atoms, the spin-flip component $\Delta E^{i}_{\uparrow \Rightarrow \downarrow}$ is quite large and has a similar value to that of the spin-preserving component $\Delta E^{i}_{\downarrow \Rightarrow \downarrow}$. This is the reason why the Bruno relation does not hold in this system as mentioned above. We can naturally understand the origin of this behavior by analyzing the LDOSs of interfacial Fe atoms shown in Fig. \ref{pert-ldos_FeMgO_a-Fe}(b). As seen from the inset of the figure, the majority-spin state has a finite $d_{3 z^2 - r^2}$ LDOS around the Fermi level. Since these $d_{3 z^2 - r^2}$ states give finite values of $\langle d_{yz},\downarrow |L^{i}_{x}| d_{3 z^2 - r^2},\uparrow \rangle$, the large positive $\Delta E^{i}_{\uparrow \Rightarrow \downarrow}$ can occur following Eq. (\ref{eq:updn}) \cite{note2}. This is the reason for the magnitude relation $\Delta E^{i}_{\downarrow \Rightarrow \downarrow} \approx \Delta E^{i}_{\uparrow \Rightarrow \downarrow}$ in this system. This feature can also be confirmed by the ${\bf k}_{\parallel}$ dependence of $\Delta E({\bf k}_{\parallel})$ and band structures of the supercell shown in Figs. \ref{FeMgO-aFe_k-resolved}(a)--\ref{FeMgO-aFe_k-resolved}(c). The major difference from the case of $a=a_{\rm MgO}/\sqrt{2}$ is that the majority-spin $d_{3z^2-r^2}$ band crosses the Fermi level around the $\Gamma$ point as shown in Fig. \ref{FeMgO-aFe_k-resolved}(b). Thus, the occupied $d_{3z^2-r^2}$ states in this band and the unoccupied $d_{yz}$ states in the minority-spin bands give finite values of $\langle d_{yz},\downarrow |L^{i}_{x}| d_{3 z^2 - r^2},\uparrow \rangle$ [see Figs. \ref{FeMgO-aFe_k-resolved}(b) and \ref{FeMgO-aFe_k-resolved}(c)], by which $\Delta E^{i}_{\uparrow \Rightarrow \downarrow}$ becomes larger compared to the case of $a=a_{\rm MgO}/\sqrt{2}$.
An experimentally realized Fe/MgO heterostructure is expected to have an intermediate in-plane lattice constant between $a_{\rm Fe}=2.866$\,{\AA} and $a_{\rm MgO}/\sqrt{2}=2.98$\,{\AA}. Although the anisotropy energy from spin-flip scattering $\Delta E^{i}_{\uparrow \Rightarrow \downarrow}$ is sensitive to the in-plane lattice constant, we can conclude from our results that Fe/MgO has a larger positive $\Delta E^{i}_{\uparrow \Rightarrow \downarrow}$ than Fe/MgAl$_2$O$_4$. This is a possible explanation for the fact that the experimentally observed $K_{\rm i}$ in Fe/MgO is larger than that in Fe/MgAl$_2$O$_4$.

\section{\label{effectofWinsertion} a way to obtain larger PMA}
In order to obtain larger interfacial PMA for MTJs with the MgAl$_2$O$_4$ barrier, we propose an insertion of thin W layers between MgAl$_2$O$_4$ and the Fe electrode as shown in Fig. \ref{W-insertion}(a). The insertion of W layers at the interface of Fe/MgAl$_2$O$_4$ is based on the theoretical prediction of huge PMA in the Fe/W(001) multilayer with the in-plane lattice constant of bulk bcc Fe \cite{2017Okamoto-APEX}. Experimentally, Matsumoto and co-workers confirmed the large change of magnetic anisotropy from negative to positive by reducing the W-layer thickness in the Fe/W(001) multilayer, where the in-plane lattice constant of W approaches from that of bulk W to that of bulk Fe \cite{2016Matsumoto-IEEEMAG,2017Okamoto-APEX}. Motivated by these theoretical and experimental results, we examined the possibility that the insertion of thin W layers into the interface of Fe/MgAl$_2$O$_4$ with the in-plane lattice constant of bcc Fe can enhance the interfacial PMA in this junction. In Fig. \ref{W-insertion}(b), we show the calculated values of $K_{\rm i}$ in Fe/W($n$)/MgAl$_2$O$_4$(001) and Fe/W($n$)/MgO(001). As in-plane lattice constants, we adopted $a=2\,a_{\rm Fe}$ for the MgAl$_2$O$_4$-based junction and $a=a_{\rm Fe}$ and $a=a_{\rm MgO}/\sqrt{2}$ for the MgO-based junction. As can be seen in Fig. \ref{W-insertion}(b), Fe/W(3--5)/MgAl$_2$O$_4$ and Fe/W(4--5)/MgO with $a=a_{\rm Fe}$ have large positive $K_{\rm i}$, indicating that the insertion of W layers significantly enhances the PMA in these junctions. Note that such a large enhancement was not obtained in very thin W cases ($n=1$--$2$). Thus, at least 3 layers of W are required for enhancing PMA. On the other hand, Fe/W($n$)/MgO with $a=a_{\rm MgO}/\sqrt{2}$ shows a negative or small $K_{\rm i}$ for any layer number of W up to $n=5$, which means that the insertion of W layers degrades the interfacial PMA in this junction because of the lattice mismatch between Fe and MgO. These results are consistent with those of Fe/W multilayers in Ref. \cite{2017Okamoto-APEX}. We can conclude that the insertion of W layers into the interface of Fe/MgAl$_2$O$_4$ is a promising way to obtain huge PMA owing to the good lattice matching between Fe and MgAl$_2$O$_4$, indicating the advantage of MgAl$_2$O$_4$ as compared with MgO. Perhaps it might be even better to use W layers as underlayers of Fe/MgAl$_2$O$_4$, because the interfacial insertion of nonmagnetic metals between FMs and insulator barriers tends to decrease TMR ratios of MTJs. However, further analysis for the PMA in such a system is beyond the scope of this study and will be addressed in our future work.

From the second-order perturbation analysis, we found that the large PMA in the Fe/W/MgAl$_2$O$_4$ multilayer is mainly attributed to perturbation processes through unoccupied majority-spin states ($\Delta E^{i}_{\downarrow \Rightarrow \uparrow}$ and $\Delta E^{i}_{\uparrow \Rightarrow \uparrow}$) of the middle-layer W atoms. Figures \ref{W-ldos}(a)--\ref{W-ldos}(d) show the projected LDOSs of middle-layer W atoms in Fe(5)/W(3)/ MgAl$_2$O$_4$(9) and Fe(5)/W(3)/MgO(5) ($a=a_{\rm Fe}$ and $a=a_{\rm MgO}/\sqrt{2}$). As can be seen in Figs. \ref{W-ldos}(a) and \ref{W-ldos}(c), there are large unoccupied $d_{xy}$ and $d_{yz}$ ($d_{zx}$) states just above the Fermi level, when the in-plane lattice constant corresponds to that of bcc Fe ($a=a_{\rm Fe}$). Because W is a transition metal element with less than half $d$ electrons, it has unoccupied majority-spin $d$ states.  These unoccupied majority-spin states provide a considerable contribution to the PMA through the second-order perturbation of the spin-orbit interaction between unoccupied majority-spin states and occupied states in both the spin channels. In the present case, the matrix elements $\langle d_{xy},\uparrow|L^{i}_z|d_{x^2-y^2},\uparrow \rangle$ and $\langle d_{yz},\uparrow|L^{i}_x|d_{3z^2-r^2},\downarrow \rangle$ show positive contributions to the PMA of Fe(5)/W(3)/MgAl$_2$O$_4$(9) and Fe(5)/W(3)/MgO(5) ($a=a_{\rm Fe}$) in the perturbation processes.

\section{summary}
We theoretically investigated interfacial magnetocrystalline anisotropy in Fe/MgAl$_2$O$_4$, which has a potential applicability to spintronic devices because of its quite small lattice mismatch at the interface. By means of density functional theory, we calculated interfacial anisotropy constant $K_{\rm i}$ of this system and compared it with those of two Fe/MgO systems with different in-plane lattice constants. We found that Fe/MgAl$_2$O$_4$ has interfacial perpendicular magnetic anisotropy (PMA) with $K_{\rm i} \approx 1.2\, {\rm mJ/m^2}$, which is slightly smaller than that of Fe/MgO systems. By carrying out second-order perturbation calculations on the PMA in combination with detailed analyses of the LDOSs and band structures, we clarified that the smaller $K_{\rm i}$ in Fe/MgAl$_2$O$_4$ is due to the smaller positive anisotropy energy from spin-flip electron scattering. We finally proposed insertion of tungsten (W) into the interface of Fe/MgAl$_2$O$_4$ as a possible way to obtain larger interfacial PMA. We showed that such insertion enhances $K_{\rm i}$ of Fe/MgAl$_2$O$_4$ to $\gtrsim 3\,{\rm mJ/m^2}$.

{\it Note added in proof.}\,\,\, Recently, Xiang {\it et al.} \cite{2018Xiang-APEX} reported detailed experimental results on the interfacial PMA in Fe/MgAl2O4, which are also consistent with our present results.

\begin{acknowledgments}
The authors are grateful to K. Hono, S. Mitani, J. Okabayashi, H. Sukegawa, M. Tsujikawa, and K. Nawa for useful discussions and critical comments. This work was partly supported by Grants-in-Aid for Scientific Research (S) (Grant No. 16H06332) and (B) (Grant No. 16H03852) from the Ministry of Education, Culture, Sports, Science and Technology, Japan, by NIMS MI$^2$I, and also by the ImPACT Program of the Council for Science, Technology and Innovation, Japan. The crystal structures of the supercells were visualized using VESTA \cite{2011Momma-JAC}.
\end{acknowledgments}


\begin{thebibliography}{99}
\bibitem{2016Dieny-Wiley} B. Dieny, R. B. Goldfarb, and K. J. Lee, {\it Introduction to Magnetic Random-access Memory} (Wiley, Hoboken, NJ, 2016).
\bibitem{2009Wu-APL} F. Wu, S. Mizukami, D. Watanabe, H. Naganuma, M. Oogane, Y. Ando, and T. Miyazaki, Appl. Phys. Lett. {\bf 94}, 122503 (2009).
\bibitem{2011Mizukami-APL} S. Mizukami, F. Wu, A. Sakuma, J. Walowski, D. Watanabe, T. Kubota, X. Zhang, H. Naganuma, M. Oogane, Y. Ando, and T. Miyazaki, Phys. Rev. Lett. {\bf 106}, 117201 (2011).
\bibitem{2012Kurt-APL} H. Kurt, N. Baadji, K. Rode, M. Venkatesan, P. S. Stamenov, S. Sanvito, and J. M. D. Coey, Appl. Phys. Lett. {\bf 101}, 132410 (2012).
\bibitem{2013Mizukami-APEX} S. Mizukami, A. Sakuma, A. Sugihara, T. Kubota, Y. Kondo, H. Tsuchiura, and T. Miyazaki, Appl. Phys. Express {\bf 6}, 123002 (2013).
\bibitem{1995Klemmer-SMM} T. Klemmer, D. Hoydick, H. Okumura, B. Zhang, and W. A. Soffa, Scr. Metall. Mater. {\bf 33}, 1793 (1995).
\bibitem{2001Butler-PRB} W. H. Butler, X.-G. Zhang, T. C. Schulthess, and J. M. MacLaren, Phys. Rev. B {\bf 63}, 054416 (2001).
\bibitem{2001Mathon-PRB} J. Mathon and A. Umerski, Phys. Rev. B {\bf 63}, 220403(R) (2001).
\bibitem{2004Parkin-NatMat} S. S. P. Parkin, C. Kaiser, A. Panchula, P. M. Rice, B. Hughes, M. Samant, and S.-H. Yang, Nat. Mater. {\bf 3}, 862 (2004).
\bibitem{2004Yuasa-NatMat} S. Yuasa, T. Nagahama, A. Fukushima, Y. Suzuki, and K. Ando, Nat. Mater. {\bf 3}, 868 (2004).
\bibitem{2010Ikeda-NatMat} S. Ikeda, K. Miura, H. Yamamoto, K. Mizunuma, H. D. Gan, M. Endo, S. Kanai, J. Hayakawa, F. Matsukura, and H. Ohno, Nat. Mater. {\bf 9}, 721 (2010).
\bibitem{2013Koo-APL} J. W. Koo, S. Mitani, T. T. Sasaki, H. Sukegawa, Z. C. Wen, T. Ohkubo, T. Niizeki, K. Inomata, and K. Hono, Appl. Phys. Lett. {\bf 103}, 192401 (2013).
\bibitem{2014Koo-JPD} J. W. Koo , H. Sukegawa , S. Kasai , Z. C. Wen , and S. Mitani , J. Phys. D: Appl. Phys. {\bf 47}, 322001 (2014).
\bibitem{2011Yang-PRB} H. X. Yang, M. Chshiev, B. Dieny, J. H. Lee, A. Manchon, and K. H. Shin, Phys. Rev. B {\bf 84}, 054401 (2011).
\bibitem{2014Zhang-PRB} J. Zhang, C. Franz, M. Czerner, and C. Heiliger, Phys. Rev. B {\bf 90}, 184409 (2014).
\bibitem{2011Wen-APL} Z. Wen, H. Sukegawa, S. Mitani, and K. Inomata, Appl. Phys. Lett. {\bf 98}, 242507 (2011).
\bibitem{2012Wen-APEX} Z. C. Wen, H. Sukegawa, S. Kasai, M. Hayashi, S. Mitani, and K. Inomata, Appl. Phys. Express {\bf 5}, 063003 (2012).
\bibitem{2016Shiota-APEX} Y. Shiota, T. Nozaki, S. Tamaru, K. Yakushiji, H. Kubota, A. Fukushima, S. Yuasa, and  Y.Suzuki, Appl. Phys. Express {\bf 9}, 013001 (2016).
\bibitem{2010Nakamura-PRB} K. Nakamura, T. Akiyama, T. Ito, M. Weinert, and A. J. Freeman, Phys. Rev. B {\bf 81}, 220409(R) (2010).
\bibitem{2013Hallal-PRB} A. Hallal, H. X. Yang, B. Dieny, and M. Chshiev, Phys. Rev. B {\bf 88}, 184423 (2013).
\bibitem{1989Bruno-PRB} P. Bruno, Phys. Rev. B {\bf 39}, 865(R) (1989).
\bibitem{2013Miura-JAP} Y. Miura, M. Tsujikawa, and M. Shirai, J. Appl. Phys. {\bf 113}, 233908 (2013).
\bibitem{2017Masuda-PRB_pma} K. Masuda, S. Kasai, Y. Miura, and K. Hono, Phys. Rev. B {\bf 96}, 174401 (2017).
\bibitem{2014Okabayashi-APL} J. Okabayashi, J. W. Koo, H. Sukegawa, S. Mitani, Y. Takagi, and T. Yokoyama, Appl. Phys. Lett. {\bf 105}, 122408 (2014).
\bibitem{2010Sukegawa-APL} H. Sukegawa, H. Xiu, T. Ohkubo, T. Furubayashi, T. Niizeki, W. Wang, S. Kasai, S. Mitani, K. Inomata, and K. Hono, Appl. Phys. Lett. {\bf 96}, 212505 (2010).
\bibitem{2012Sukegawa-PRB} H. Sukegawa, Y. Miura, S. Muramoto, S. Mitani, T. Niizeki, T. Ohkubo, K. Abe, M. Shirai, K. Inomata, and K. Hono, Phys. Rev. B {\bf 86}, 184401 (2012).
\bibitem{2016Belmoubarik-APL} M. Belmoubarik, H. Sukegawa, T. Ohkubo, S. Mitani, and K. Hono, Appl. Phys. Lett. {\bf 108}, 132404 (2016).
\bibitem{2016Scheike-APEX} T. Scheike, H. Sukegawa, K. Inomata, T. Ohkubo, K. Hono, and S. Mitani, Appl. Phys. Express {\bf 9}, 053004 (2016).
\bibitem{2014Koo-PSS} J. Koo, H. Sukegawa, and S. Mitani, Phys. Status Solidi RRL {\bf 8}, 841 (2014).
\bibitem{2014Tao-APL} B. S. Tao, D. L. Li, Z. H. Yuan, H. F. Liu, S. S. Ali, J. F. Feng, H. X. Wei, X. F. Han, Y. Liu, Y. G. Zhao, Q. Zhang, Z. B. Guo, and X. X. Zhang, Appl. Phys. Lett. {\bf 105}, 102407 (2014).
\bibitem{2017Sukegawa-APL} H. Sukegawa, J. P. Hadorn, Z. Wen, T. Ohkubo, S. Mitani, and K. Hono, Appl. Phys. Lett. {\bf 110}, 112403 (2017).
\bibitem{1996Kresse-PRB} G. Kresse and J. Furthm\"uller, Phys. Rev. B {\bf 54}, 11169 (1996).
\bibitem{1996Perdew-PRL} J. P. Perdew, K. Burke, and M. Ernzerhof, Phys. Rev. Lett. {\bf 77}, 3865 (1996).
\bibitem{1994Bloechl-PRB} P. E. Bl\"ochl, Phys. Rev. B {\bf 50}, 17953 (1994).
\bibitem{1999Kresse-PRB} G. Kresse and D. Joubert, Phys. Rev. B {\bf 59}, 1758 (1999).
\bibitem{remark_interface} One of the present authors found this fact in the case of Fe/MgAl$_2$O$_4$ \cite{2012Miura-PRB}. In the case of Fe/MgO, this fact was confirmed both experimentally \cite{1988Urano-JPSJ} and theoretically \cite{1991Li-PRB}.
\bibitem{2012Miura-PRB} Y. Miura, S. Muramoto, K. Abe, and M. Shirai, Phys. Rev. B {\bf 86}, 024426 (2012).
\bibitem{1988Urano-JPSJ} T. Urano and T. Kanaji, J. Phys. Soc. Jpn. {\bf 57}, 3403 (1988).
\bibitem{1991Li-PRB} C. Li and A. J. Freeman, Phys. Rev. B {\bf 43}, 780 (1991).
\bibitem{2017Masuda-PRB_bias} K. Masuda and Y. Miura, Phys. Rev. B {\bf 96}, 054428 (2017).
\bibitem{1990Daalderop-PRB} G. H. O. Daalderop, P. J. Kelly, and M. F. H. Schuurmans, Phys. Rev. B {\bf 41}, 11919 (1990).
\bibitem{remark_kpoints} Note that the required number of in-plane k-point in Fe/MgAl$_2$O$_4$ is about half that of Fe/MgO, because the in-plane lattice constant of Fe/MgAl$_2$O$_4$ is about twice as that of Fe/MgO.
\bibitem{2000Grzybowski-PRB} A. Grzybowski, E. Gw{\'o}{\'z}d{\'z}, and A. Br{\'o}dka, Phys. Rev. B {\bf 61}, 6706 (2000).
\bibitem{2013Miura-JPCM} Y. Miura, S. Ozaki, Y. Kuwahara, M. Tsujikawa, K. Abe, and M. Shirai, J. Phys. Condens. Matter {\bf 25}, 106005 (2013).
\bibitem{2007Andersson-PRL} C. Andersson, B. Sanyal, O. Eriksson, L. Nordstr\"om, O. Karis, D. Arvanitis, T. Konishi, E. Holub-Krappe, and J. H. Dunn, Phys. Rev. Lett. {\bf 99}, 177207 (2007).
\bibitem{1993Wang-PRB} D. S. Wang, R. Wu, and A. J. Freeman, Phys. Rev. B {\bf 47}, 14932 (1993).
\bibitem{1998Laan-JPCM} G. van der Laan, J. Phys.: Condens. Matter {\bf 10}, 3239 (1998).
\bibitem{2009Nakamura-PRL} K. Nakamura, R. Shimabukuro, Y. Fujiwara, T. Akiyama, T. Ito, and A. J. Freeman, Phys. Rev. Lett. {\bf 102}, 187201 (2009).
\bibitem{1994Daalderop-PRB} G. H. O. Daalderop, P. J. Kelly, and M. F. H. Schuurmans, Phys. Rev. B {\bf 50}, 9989 (1994).
\bibitem{2012Ouazi-NatCom} S. Ouazi, S. Vlaic, S. Rusponi, G. Moulas, P. Buluschek, K. Halleux, S. Bornemann, S. Mankovsky, J. Min{\'a}r, J. B. Staunton, H. Ebert, and H. Brune, Nat. Commun. {\bf 3}, 1313 (2012).
\bibitem{note1} As mentioned in Sec. \ref{methods}, the in-plane lattice constant of Fe/MgAl$_2$O$_4$ supercell is approximately twice as those of Fe/MgO supercells. Thus, we plotted $\Delta E({\rm k}_{\parallel})/4$ for reasonable comparison with Fe/MgO cases shown in Figs. \ref{FeMgO-aMgO_k-resolved}(a) and \ref{FeMgO-aFe_k-resolved}(a).
\bibitem{note2} Note that the small values of the energy difference $\epsilon_{u_{\downarrow}}-\epsilon_{o_{\uparrow}}$ between the unoccupied $d_{yz}$ ($d_{zx}$) and occupied $d_{3 z^2 - r^2}$ states enhances the value of $\Delta E^{i}_{\uparrow \Rightarrow \downarrow}$.
\bibitem{2017Okamoto-APEX} Y. Matsumoto, Y. Miura, S. Okamoto, N. Kikuchi, and O. Kitakami, Appl. Phys. Express {\bf 10}, 063005 (2017).
\bibitem{2016Matsumoto-IEEEMAG} Y. Matsumoto, S. Okamoto, N. Kikuchi, O. Kitakami, Y. Miura, M. Suzuki, M. Mizumaki, and N. Kawamura, IEEE Trans. Magn. {\bf 51}, 1 (2015).
\bibitem{2018Xiang-APEX} Q. Xiang, R. Mandal, H. Sukegawa, Y. K. Takahashi, and S. Mitani, Appl. Phys. Express {\bf 11}, 063008 (2018).
\bibitem{2011Momma-JAC} K. Momma and F. Izumi, J. Appl. Cryst. {\bf 44}, 1272 (2011).
\end{thebibliography}

\newpage

\begin{figure}
\includegraphics[width=12.0cm]{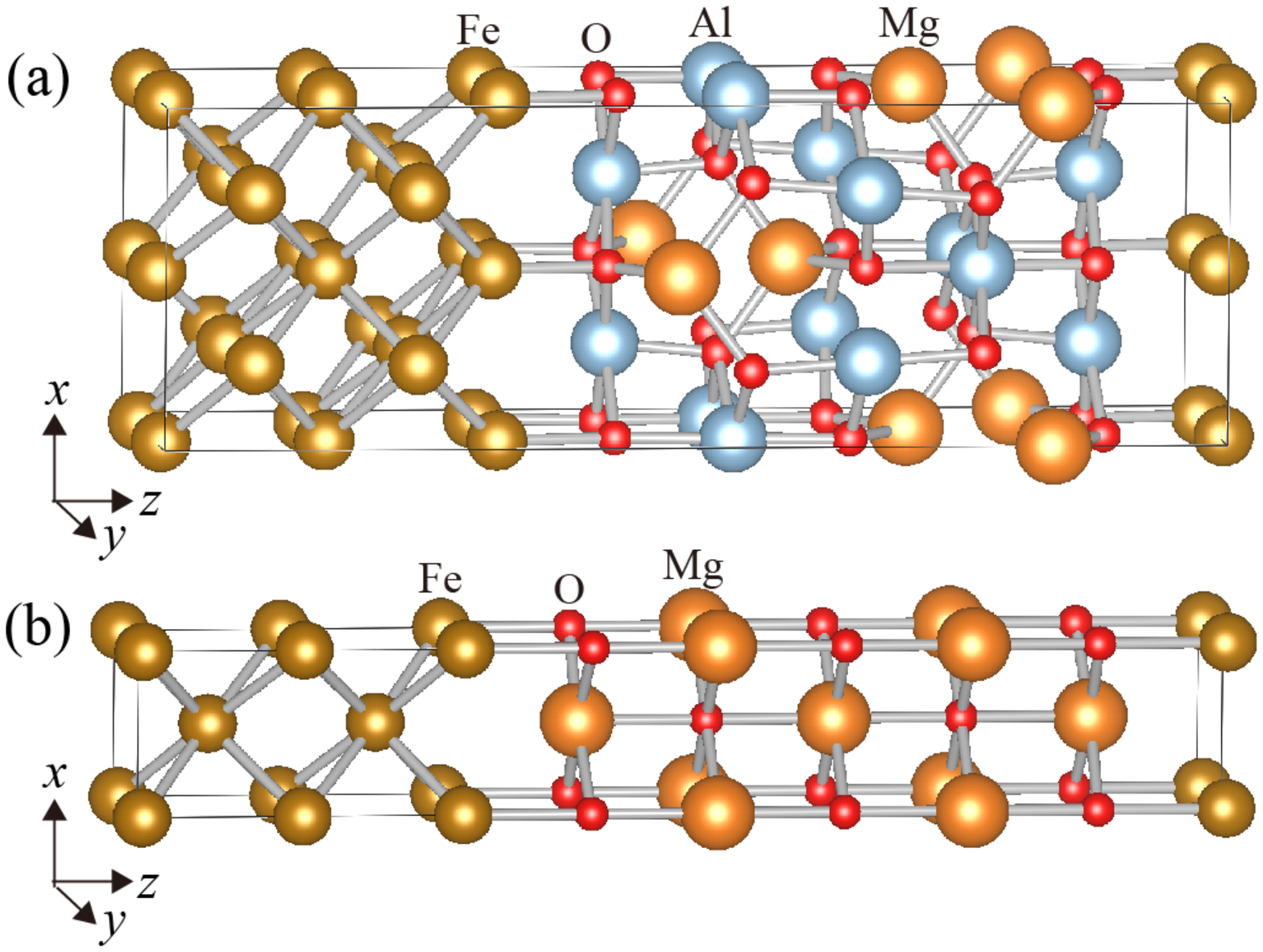}
\caption{\label{supercells} Supercells of (a) Fe(5)/MgAl$_2$O$_4$(9) and (b) Fe(5)/MgO(5).}
\end{figure}

\begin{figure}
\includegraphics[width=15.0cm]{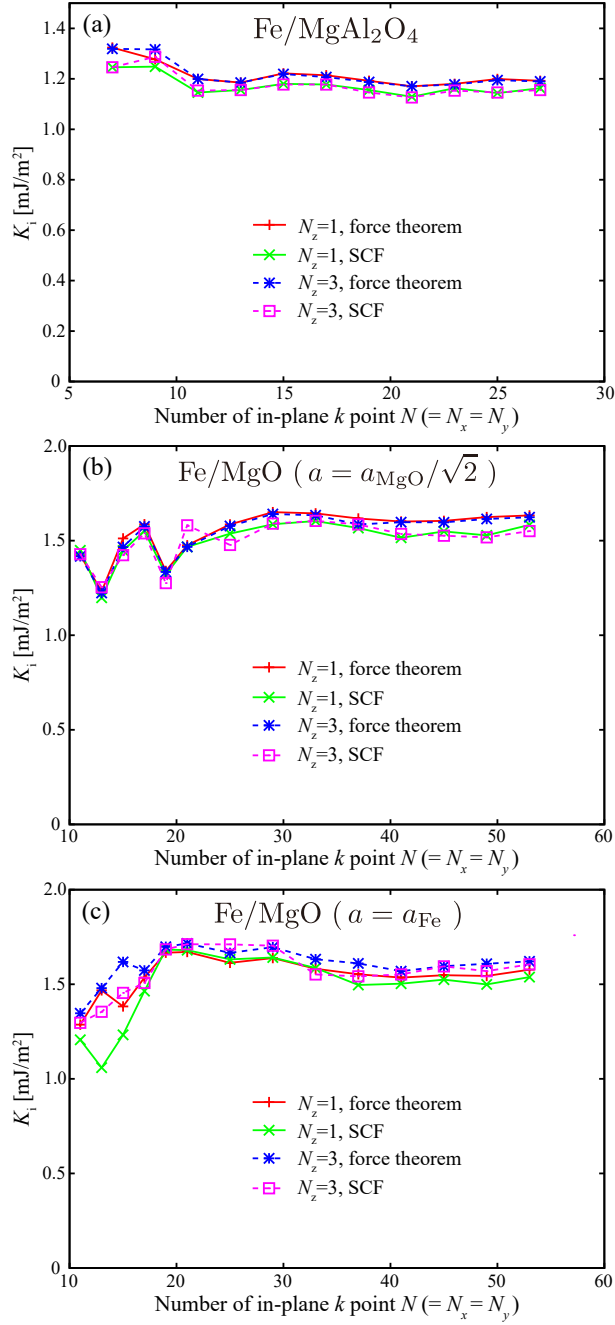}
\caption{\label{kdep} The interfacial anisotropy constant $K_{\rm i}$ as a function of the number of in-plane k-point $N$ ($=N_x=N_y$) in (a) Fe/MgAl$_2$O$_4$, (b) Fe/MgO ($a=a_{\rm MgO}/\sqrt{2}$), and (c) Fe/MgO ($a=a_{\rm Fe}$).}
\end{figure}

\begin{table}
\caption{\label{tab1}
List of $K_{\rm i}$, $E_{\rm demag}\,t$, $K_{\rm eff}\,t$, $\Delta M_{\rm orb,i}$, and $M_{\rm spin,i}$ obtained in this study.
}
\begin{ruledtabular}
\begin{tabular}{lccccc}
\textrm{\scriptsize System}&
\textrm{\scriptsize $K_{\rm i}$\,(${\rm mJ/m^2}$)}&
\textrm{\scriptsize $E_{\rm demag}\,t$\,(${\rm mJ/m^2}$)}&
\textrm{\scriptsize $K_{\rm eff}\,t$\,(${\rm mJ/m^2}$)}&
\textrm{\scriptsize $\Delta M_{\rm orb,i}$\,($\mu_{\rm B}$/atom)}&
\textrm{\scriptsize $M_{\rm spin,i}$\,($\mu_{\rm B}$/atom)}\\
\colrule
{\scriptsize Fe/MgAl$_2$O$_4$}    & 1.192 & -0.895 & 0.296 & 0.026 & 2.81\\
{\scriptsize Fe/MgO ($a=a_{\rm MgO}/\sqrt{2}$)} & 1.617 & -0.828 & 0.788 & 0.030 & 2.73\\
{\scriptsize Fe/MgO ($a=a_{\rm Fe}$)} & 1.552 & -0.908 & 0.643 & 0.020 & 2.78\\
\end{tabular}
\end{ruledtabular}
\end{table}

\begin{figure}
\includegraphics[width=12.0cm]{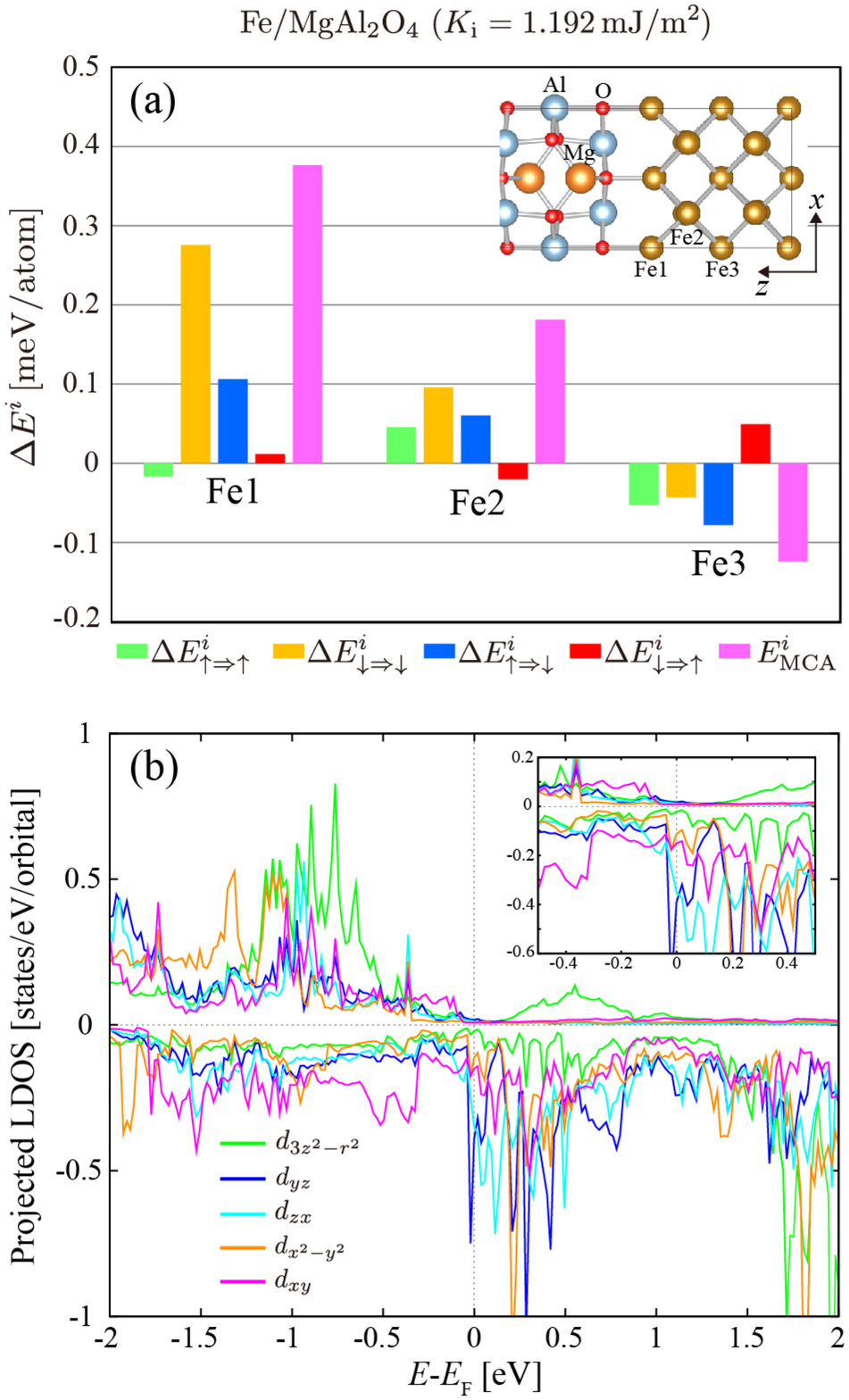}
\caption{\label{pert-ldos_FeMAO} (a) Results of second-order perturbation analysis on the interfacial PMA in Fe/MgAl$_2$O$_4$. The vertical green, yellow, blue, red, and pink bars show the values of $\Delta E^{i}_{\uparrow \Rightarrow \uparrow}$, $\Delta E^{i}_{\downarrow \Rightarrow \downarrow}$, $\Delta E^{i}_{\uparrow \Rightarrow \downarrow}$, $\Delta E^{i}_{\downarrow \Rightarrow \uparrow}$, and the local anisotropy energy $E^{i}_{\rm MCA}$, respectively, at each Fe layer. [See Eq. (\ref{eq:E2MCA_decomp}) and the corresponding text for details.] (b) Projected LDOSs for Fe $3d$ states at the interface of Fe/MgAl$_2$O$_4$. In panel (b), positive and negative values indicate the majority- and minority-spin projected LDOSs, respectively. The inset of panel (b) shows a magnified view near the Fermi level.}
\end{figure}

\begin{figure}
\includegraphics[width=12.0cm]{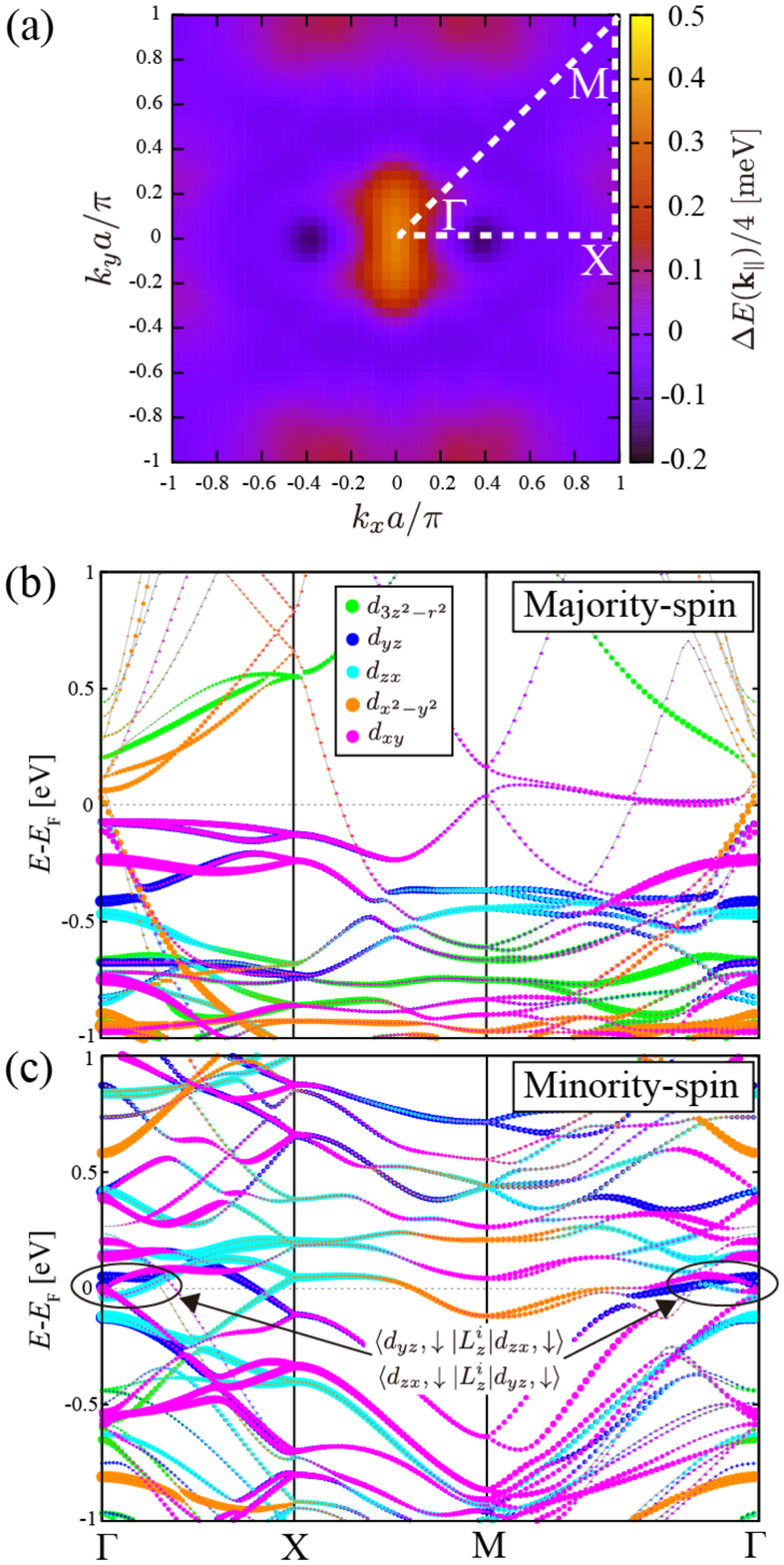}
\caption{\label{FeMAO_k-resolved} The wave-vector-resolved information on the interfacial PMA in Fe/MgAl$_2$O$_4$. (a) The in-plane wave-vector (${\bf k}_{\parallel}$) dependence of $\Delta E({\bf k}_{\parallel}) \equiv E_{[100]}({\bf k}_{\parallel})-E_{[001]}({\bf k}_{\parallel})$. (b) and (c) The band structure of the supercell in the majority- and minority-spin states, respectively. In panels (b) and (c), orbital components of each band are indicated by colors.}
\end{figure}

\begin{figure}
\includegraphics[width=12.0cm]{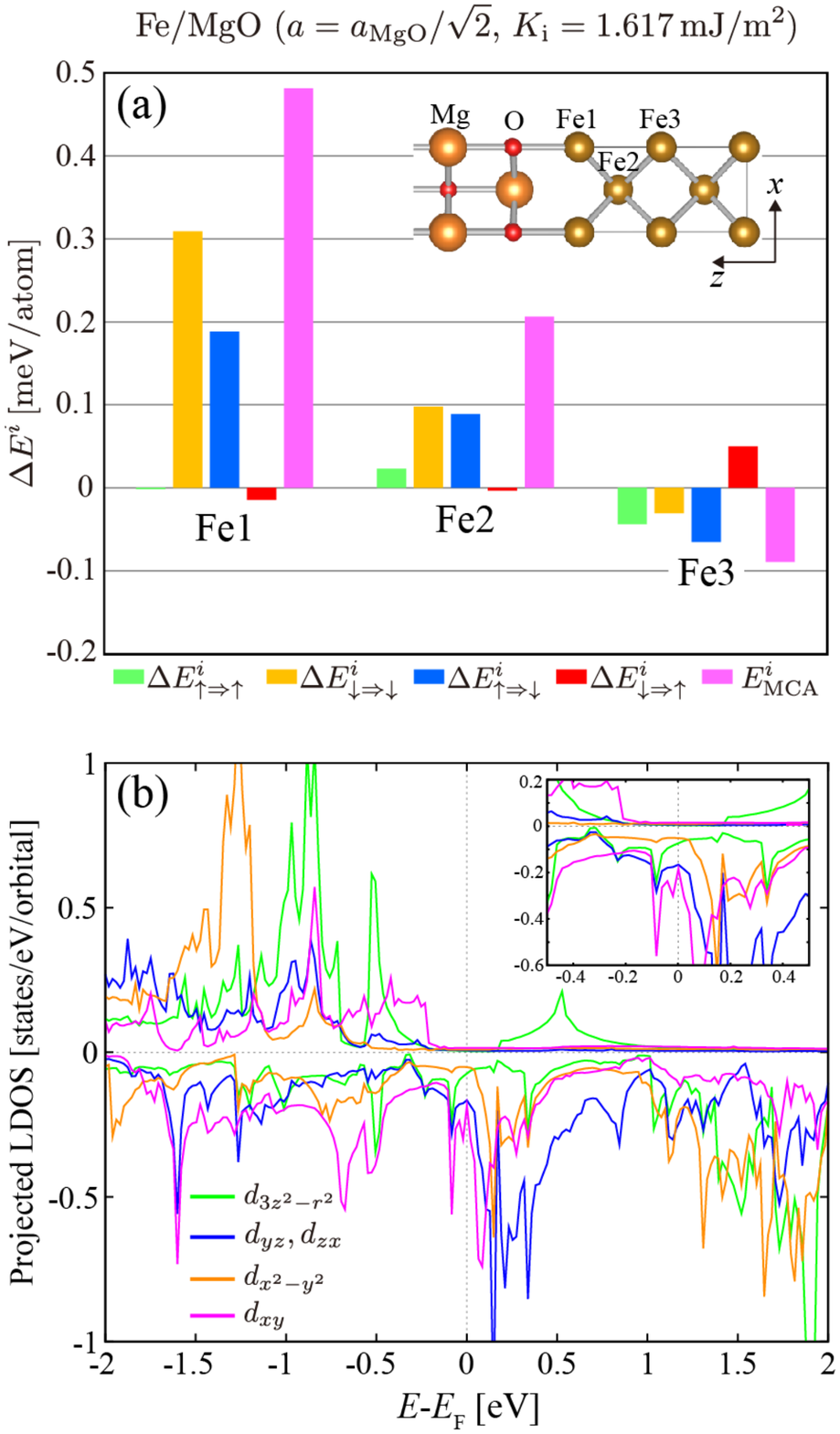}
\caption{\label{pert-ldos_FeMgO_a-MgO} The same as Fig. \ref{pert-ldos_FeMAO}, but for Fe/MgO with $a=a_{\rm MgO}/\sqrt{2}$.}
\end{figure}

\begin{figure}
\includegraphics[width=12.0cm]{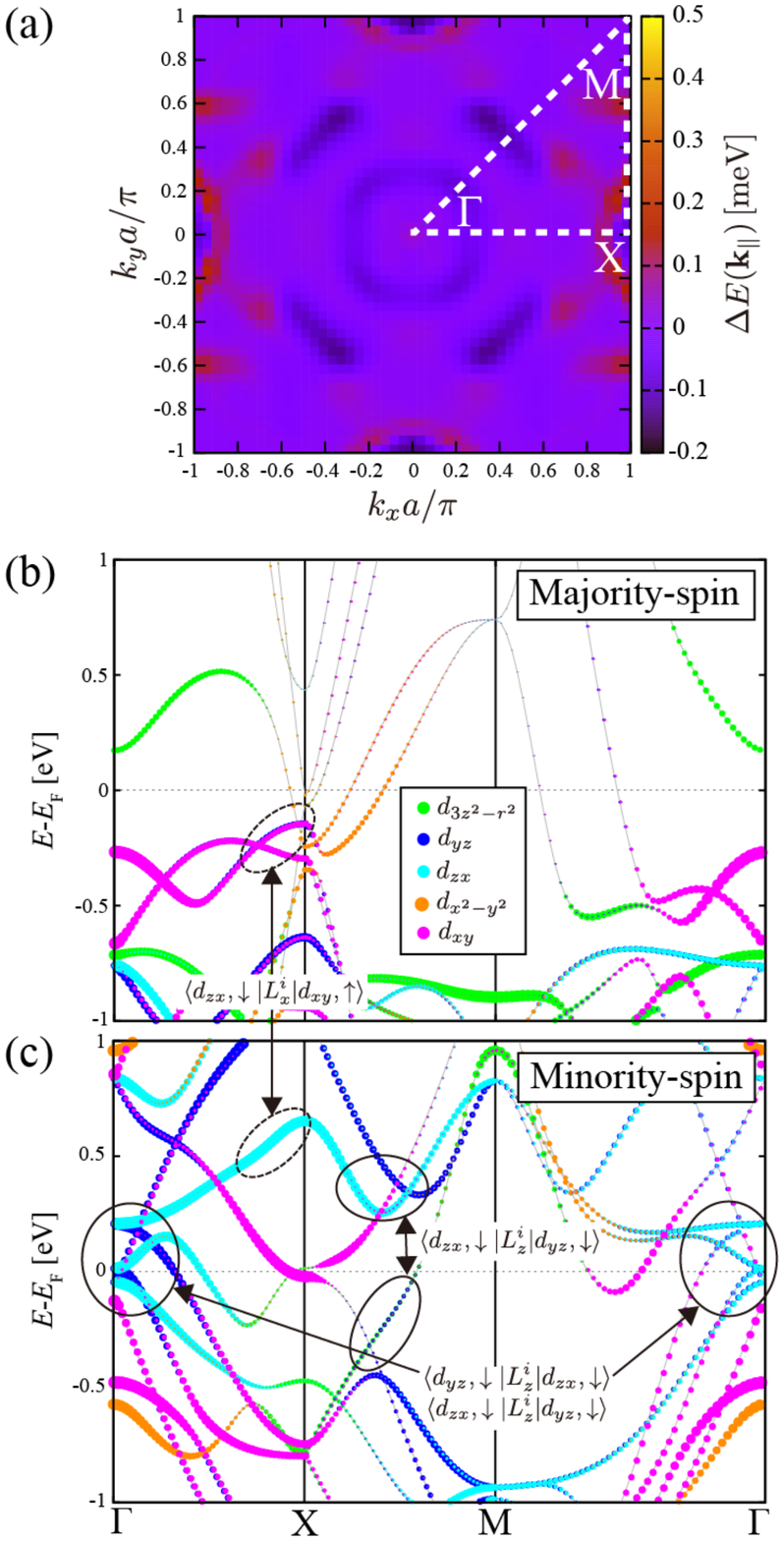}
\caption{\label{FeMgO-aMgO_k-resolved} The same as Fig. \ref{FeMAO_k-resolved}, but for Fe/MgO with $a=a_{\rm MgO}/\sqrt{2}$.}
\end{figure}

\begin{figure}
\includegraphics[width=12.0cm]{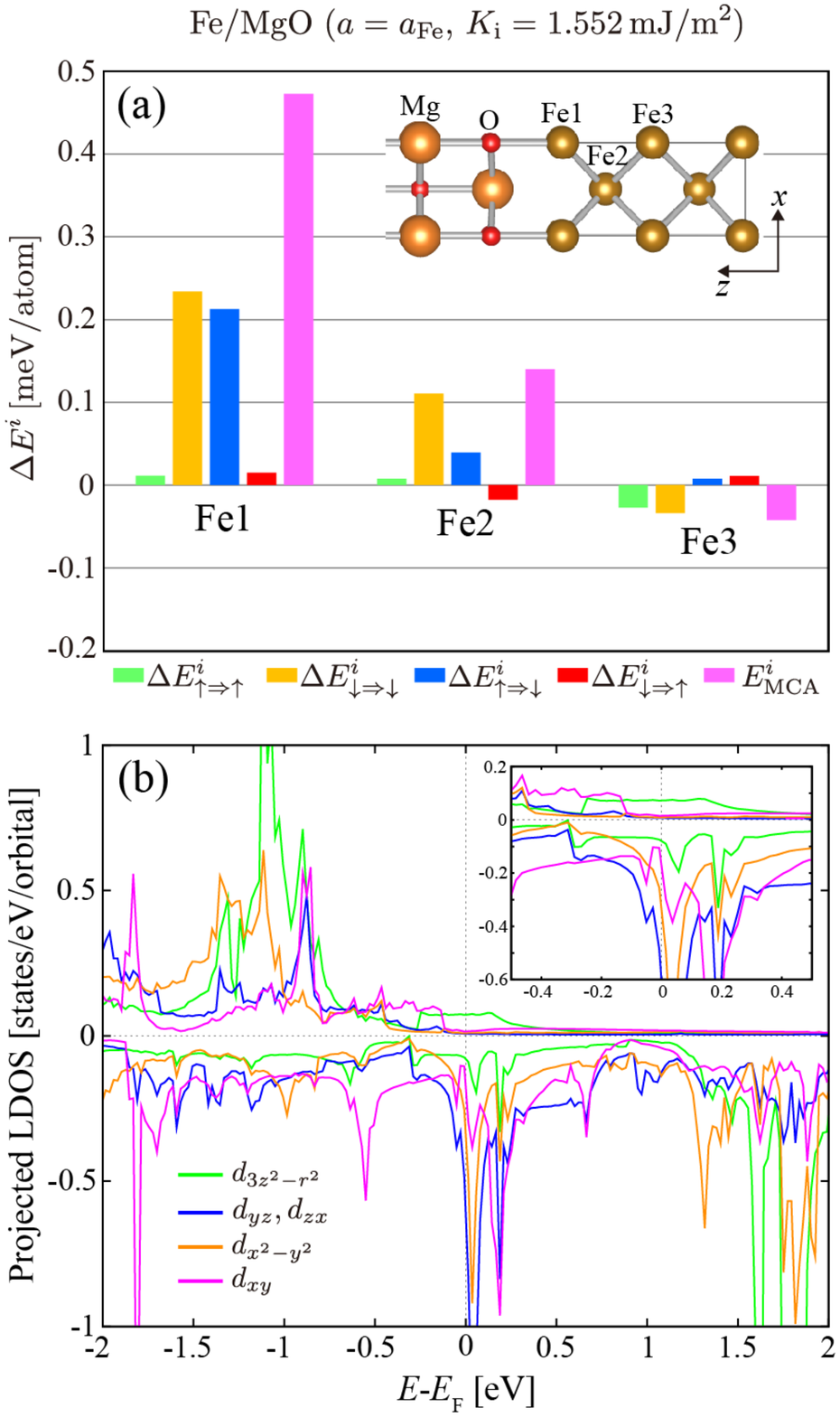}
\caption{\label{pert-ldos_FeMgO_a-Fe} The same as Fig. \ref{pert-ldos_FeMAO}, but for Fe/MgO with $a=a_{\rm Fe}$.}
\end{figure}

\begin{figure}
\includegraphics[width=12.0cm]{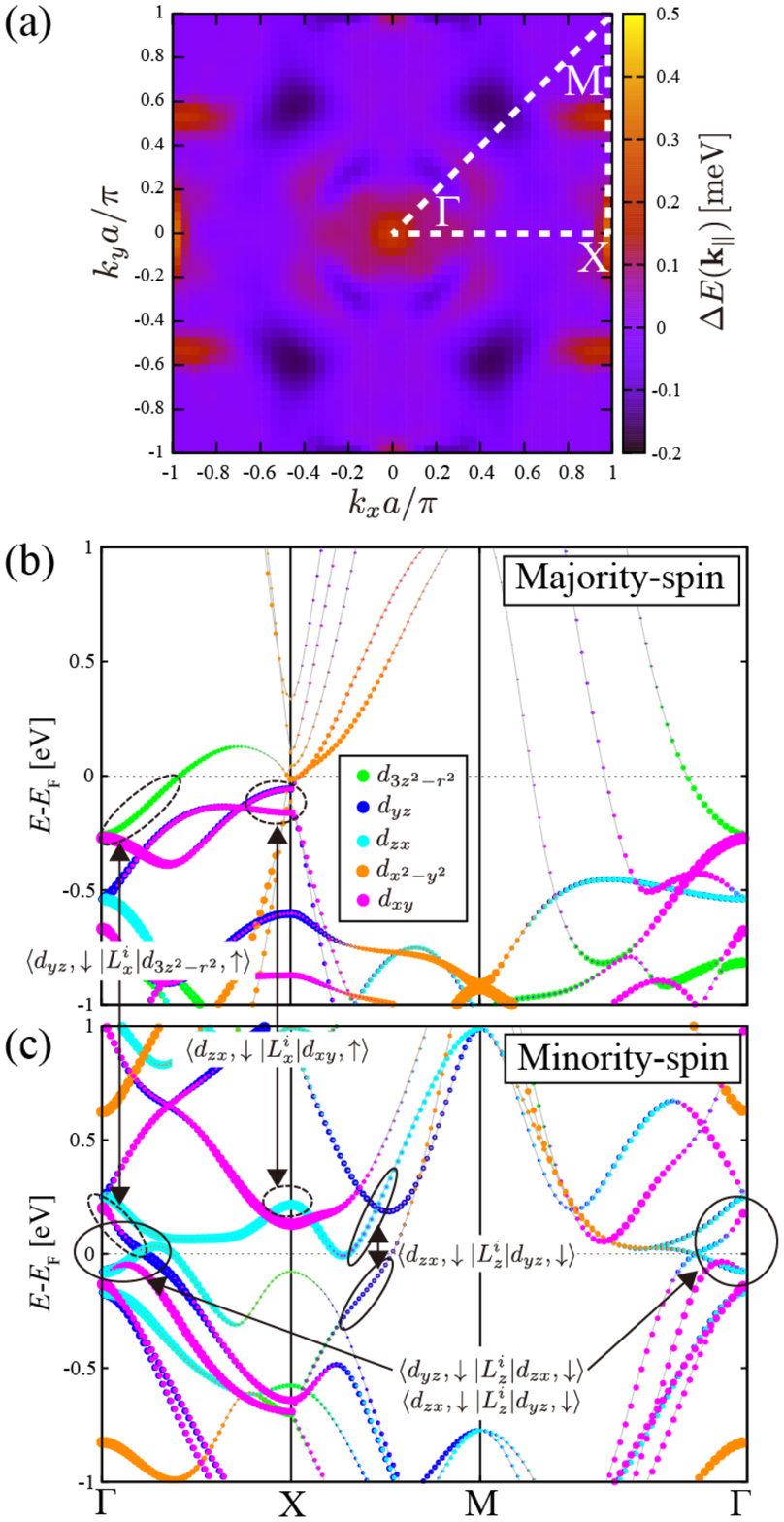}
\caption{\label{FeMgO-aFe_k-resolved} The same as Fig. \ref{FeMAO_k-resolved}, but for Fe/MgO with $a=a_{\rm Fe}$.}
\end{figure}

\begin{figure}
\includegraphics[width=13.0cm]{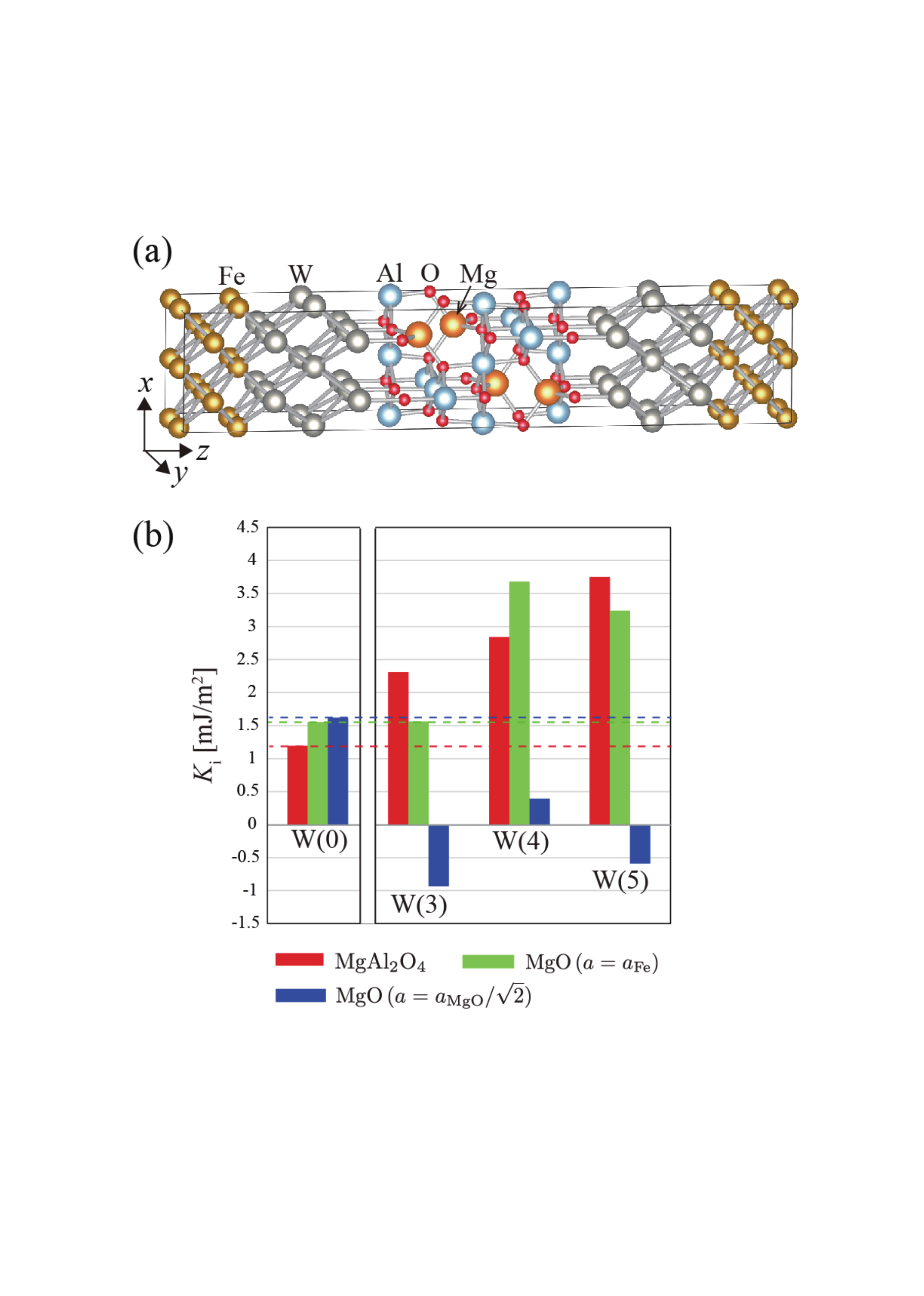}
\caption{\label{W-insertion} (a) The supercell of Fe(5)/W(3)/MgAl$_2$O$_4$(9) used in our calculation. (b) Values of $K_{\rm i}$ obtained for Fe(5)/W(3--5)/MgAl$_2$O$_4$(9) and two types of Fe(5)/W(3--5)/MgO(5) with different in-plane lattice constants. The data indicated by W(0) are those when W layers are not inserted (the values of $K_{\rm i}$ were already shown in Table \ref{tab1}).}
\end{figure}

\begin{figure}
\includegraphics[width=12.0cm]{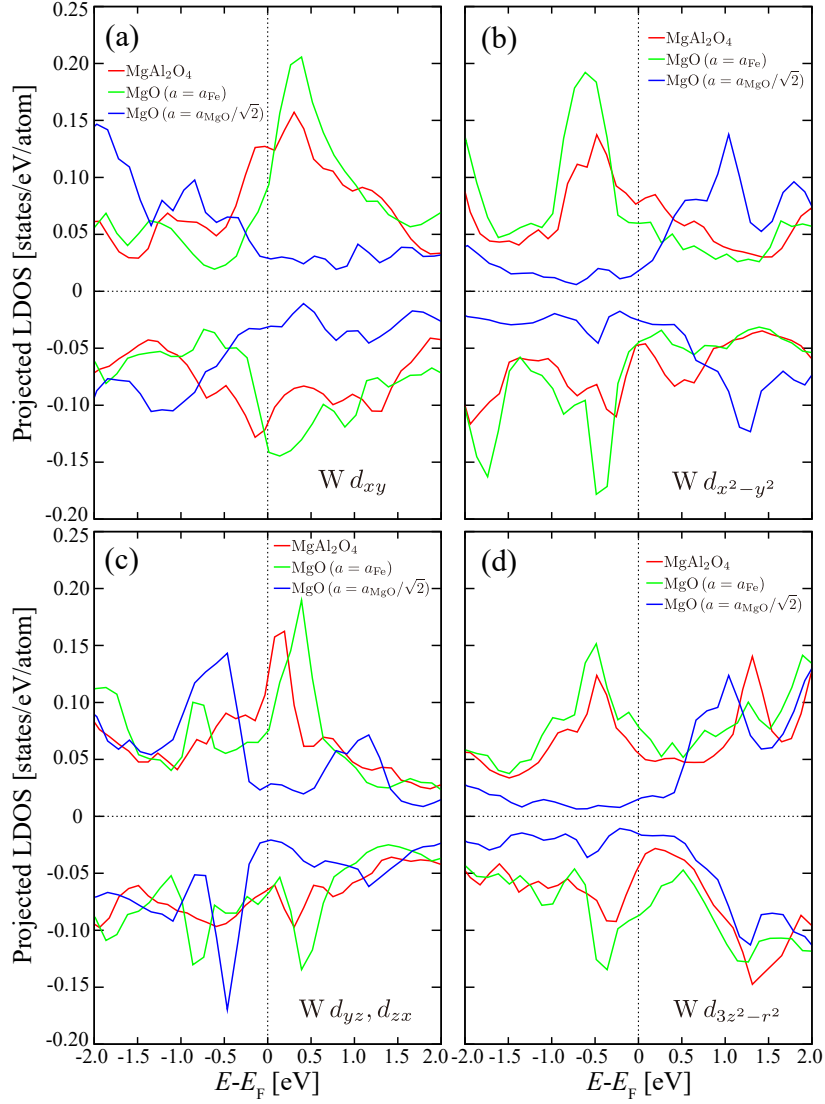}
\caption{\label{W-ldos} (a)--(d) Projected LDOSs for 3$d$ states at the middle layer of the W atom in Fe(5)/W(3)/MgAl$_2$O$_4$(9), Fe(5)/W(3)/MgO(5) ($a=a_{\rm Fe}$), and Fe(5)/W(3)/MgO(5) ($a=a_{\rm MgO}/\sqrt{2}$), where positive and negative values indicate the majority- and minority-spin projected LDOSs, respectively.}
\end{figure}

\end{document}